\documentclass[11pt]{amsart}

\usepackage[verbose=true]{microtype}
\usepackage{amsmath}
\usepackage{amssymb}
\usepackage{amsthm}
\usepackage{bbm}
\usepackage{amsrefs}
\usepackage{mathtools}
\usepackage{color}
\usepackage[colorlinks,linkcolor=black,citecolor=black,linktocpage,hypertexnames=true,pdfpagelabels]{hyperref}
\usepackage[capitalise]{cleveref}
\usepackage[all]{xy}
\usepackage{subcaption}
\usepackage{array}
\usepackage{enumerate}
\usepackage{enumitem}
\usepackage{mdframed}
\usepackage[foot]{amsaddr}
\usepackage{dsfont}

\usepackage{afterpage,longtable}

\usepackage{tikz}
\usepackage{relsize}
\usepackage{diagbox}
\usepackage{tikz-cd} 

\usepackage{multirow}

\usepackage{pifont}

\let\originalleft\left
\let\originalright\right
\renewcommand{\left}{\mathopen{}\mathclose\bgroup\originalleft}
\renewcommand{\right}{\aftergroup\egroup\originalright}
\renewcommand{\epsilon}{\varepsilon}
\newcommand{\N}{\mathbb{N}}
\newcommand{\C}{\mathbb{C}}
\newcommand{\R}{\mathbb{R}}

\newcommand{\mc}{\mathcal}

\newcommand{\tr}{\mathrm{tr}}

\newcommand{\be}{\begin{eqnarray}}
\newcommand{\ee}{\end{eqnarray}}
\newcommand{\ben}{\begin{enumerate}}
\newcommand{\een}{\end{enumerate}}
\newcommand{\bi}{\begin{itemize}}
\newcommand{\ei}{\end{itemize}}

\newcommand{\ba}{\begin{array}}
\newcommand{\ea}{\end{array}}

\newtheorem{theorem}{Theorem}

\theoremstyle{definition}

\usepackage{pgfornament}
\newcommand{\deco}{\raisebox{.6ex}{\pgfornament[width=.5cm,color = darkgray]{3}}}

\begin{document}

\title[One wonderland through two looking glasses]{\normalsize Quantum information theory and  \\  \vspace{2mm} 
free semialgebraic 
geometry:   \\ \vspace{2mm}
one wonderland through two looking  \\ \vspace{2mm}
glasses}
\author{Gemma De las Cuevas}
\address{Institut f\"ur Theoretische Physik, Technikerstr.\ 21a, A6020 Innsbruck, Austria}
\author{Tim Netzer}
\address{Institut f\"ur Mathematik, Technikerstr.\ 13, A6020 Innsbruck, Austria}
\date{\today}   

\maketitle

\begin{abstract} 
We  illustrate how quantum information theory and free (i.e.\ noncommutative) semialgebraic geometry often study similar objects from different perspectives. We give examples in the context of positivity and separability, quantum magic squares, quantum correlations in non-local games, and positivity in tensor networks, and we show the benefits of combining the two perspectives. This paper is an  invitation to consider the intersection of the two fields, and should be accessible for researchers from either field. 
\end{abstract}

\bigskip
 \begin{flushright}
\scriptsize{{\em Live free or die.}} \\ 
\scriptsize{Motto of New Hampshire}
\end{flushright}

\medskip

\section{Introduction}

The ties between physics, computer science and mathematics are historically strong and multidimensional.  
It has often happened that mathematical inventions which were mere products of imagination (and thus thought to be useless for applications) have later played a crucial role in physics or computer science. 
A superb example is that of imaginary numbers  and their use in complex Hilbert spaces  in quantum mechanics.\footnote{Who would have thought that the square root of $-1$ would have any physical relevance? See \cite{Re21}.} 
Other examples include number theory and its use in cryptography, or Riemannian geometry and its role in General Relativity. 
It is also true that physicists tend to be only aware of the mathematical tools useful for them --- so there are many branches of mathematics which have not found an outlet in physics.

The relevance  of a statement depends on the glass through which we look at it. 
There are statements which are mathematically unimpressive but  physically very impressive. 
A good example is entanglement. 
Mathematically, the statement that the positivity cone of the tensor product space is larger than the tensor product of the local cones  is interesting, but not particularly wild or surprising.
Yet, the physical existence of entangled particles is, from our perspective, truly remarkable. 
In other words, while the mathematics is easy to understand, the physics is mind-blowing. 
This is particularly true regarding Bell's Theorem: 
while it is mathematically not specially deep, 
we regard the experimental violation of Bell inequalities \cite{Gi15,He15,Sh15}   as very deep indeed. 
Another example is the no-cloning theorem --- it is mathematically trivial, 
yet it has very far-reaching physical consequences.  
On the other hand, there are many mathematically impressive statements which are --- so far --- physically irrelevant. 
Finally, there are statements which can be both mathematically deep and central for quantum information theory, such as Stinespring's Dilation Theorem.

\begin{figure}[t]
\includegraphics[width=.4\columnwidth]{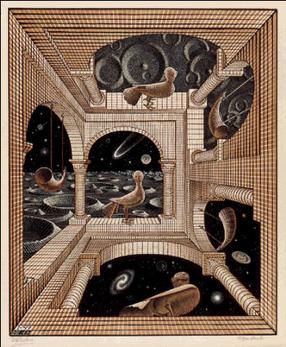}
\caption{Free semialgebraic geometry and quantum information often look at similar landscapes from different perspectives, as in the fantastic world of this woodcut by M. C. Escher. }
\label{fig:anotherworld}
\end{figure}

The goal of this paper is to illustrate how two relatively new disciplines in physics  and mathematics 
--- quantum information theory and free semialgebraic geometry --- have a lot in common (\cref{fig:anotherworld}). 
`Free' means noncommutative, because it is \emph{free of the commutation relation}. 
So free semialgebraic geometry studies noncommutative versions of semialgebraic sets. 
On the other hand, quantum information theory is (mathematically) a noncommutative generalisation of classical information theory. 
So, intuitively,  `free' is naturally linked to  `quantum'. 
Moreover, in both fields, positivity plays a very important role. 
Semialgebraic geometry examines questions  arising  from nonnegativity, like polynomial inequalities. 
In quantum information theory, quantum states are represented by positive semidefinite matrices. 
Positivity also gives rise to convexity, which is central in both fields, as we will see. 

So the two disciplines often study the same mathematical objects from different perspectives. As a consequence, they often ask different questions. 
For example, in quantum information theory, given an element of a tensor product space, one wants to know whether it is  positive semidefinite, and how this can be efficiently represented and manipulated.  
In free semialgebraic geometry, the attention is focused on the geometry of the set of all such elements (see  \cref{tab:comparison}).

\begin{longtable}[t]{| p{.47\textwidth} | p{.47\textwidth} |}
\hline
 Quantum information theory  & Free semialgebraic geometry   \vspace{1mm} 
 \\
 \emph{Emphasis on the element} & 
 \emph{Emphasis on the set }
   \vspace{1mm}
 \\ 
 \hline  \hline 
 Given $\rho=\sum_\alpha A_\alpha\otimes B_\alpha$,  is it positive semidefinite (psd)? 
 &
 Given $\{A_\alpha\}$, characterise the set of $\{B_\alpha\}$ such that $\sum_\alpha A_\alpha\otimes B_\alpha$  is psd. 
\vspace{-2mm} \\ 
 Block positive matrices /  Separable psd matrices.   
 &   Largest / smallest operator system  over the psd cone.  
   \vspace{2mm} 
   \\   
  Every POVM can be dilated to a PVM. 
& The free convex hull of the set of   PVMs is the set of POVMs.  
  \vspace{2mm} \\   
 Can a correlation matrix $p$  be realised by a  quantum strategy? 
&   Is $p$ in the free convex hull  of the free independence model?     
  \\  \hline 
\caption{ Examples of the different approaches of quantum information theory and free semialgebraic geometry in studying  essentially the same mathematical objects. The various notions will be explained throughout the paper. }
\label{tab:comparison}
\end{longtable}

We believe that much is to be learnt by bridging the gap among the two communities  --- in knowledge, notation and perspective. In this paper we hope to illustrate this point.  
 This paper is thus meant to be accessible for physicists and mathematicians.

Obviously we are not the first or the only ones to notice these similarities.
In this paper, however, we will mainly concentrate on what we have learnt from collaborating in recent years, and we will review a few other works.  
The selection of other works is not comprehensive and reflects our partial knowledge.
We also remark that quantum information theory and free semialgebraic geometry   are not the only ones studying positivity in tensor product spaces. 
\emph{Compositional distributional semantics}, for example, represents the meaning of words by positive semidefinite matrices, and the composition of meanings is thus given by positivity preserving maps --- see e.g.\ \cite{De20c,Co20}.

This article is organised as follows. 
We will first explain basic concepts in quantum information theory and free semialgebraic geometry (\cref{sec:basic}) --- the reader familiar with them can skip the corresponding section. 
Then we will explain how they are related (\cref{sec:results}), and 
we will end with some closing words (\cref{sec:concl}).

\section{Some basic concepts}
\label{sec:basic}

Here we present some basic concepts in quantum information theory (\cref{ssec:qi})
 and  free semialgebraic geometry (\cref{ssec:fsg}).

Throughout the paper we denote the set of $r\times s$ complex matrices by ${\rm Mat}_{r,s}$,  
the set of 
$r\times r$ complex  matrices by ${\rm Mat}_{r}$, and we use the identification of  ${\rm Mat}_{r}\otimes {\rm Mat}_{s}$  with ${\rm Mat}_{rs}$. 
We will also often use the real subspace of ${\rm Mat}_{r}$ containing the Hermitian elements, called ${\rm Her}_{r}$, and use that ${\rm Her}_{r}\otimes {\rm Her}_{s}$ is identified with ${\rm Her}_{rs}$. 
The $d$-fold cartesian product of ${\rm Her}_{r}$ is denoted ${\rm Her}_{r}^d$.

\subsection{Basic concepts from quantum information theory} 
\label{ssec:qi}
Here we briefly introduce some concepts from quantum information theory. 
We focus on finite-dimensional quantum systems of which we do not assume to have perfect knowledge (in the language of quantum information, these are called \emph{mixed states}). 
See, e.g.\ \cite{wi,Ni00}, for a more general overview. 

The {\it state} of a quantum system is  modelled  by a {\it normalized positive semidefinite matrix}, i.e.\ a 
$$\rho\in {\rm Mat}_d\mbox{ with }  \rho\succcurlyeq 0 \mbox{  and }  \tr(\rho)=1, 
$$ 
where $\succcurlyeq 0$  denotes positive semidefinite (psd), i.e.\ Hermitian with nonnegative eigenvalues, and the trace, $\tr$, is the sum of the diagonal elements. 
We reserve the symbol $\geqslant 0$  for nonnegative  numbers. 
A {\it measurement} on the system is modelled by a \emph{positive operator valued measure} (POVM), i.e.\ a set of psd matrices $\tau_i$ that sum to the identity: 
$$
\tau_1,\ldots,\tau_n  \in {\rm Mat}_d \mbox{ with all } \tau_i \succcurlyeq 0  \mbox{ and }  \sum_i \tau_i=I_d.$$
The probability to obtain outcome $i$ on state $\rho$ is given by 
\be\label{eq:povm}
\tr(\rho \tau_i). 
\ee
Note that these probabilities sum to 1 because of the normalisation condition on the $\tau_i$'s and $\rho$.

When the system is composed of several subsystems, 
the global state space is modelled as a tensor product of the local spaces, 
\be \label{eq:tensprod}
{\rm Mat}_d= {\rm Mat}_{d_1}\otimes \cdots \otimes {\rm Mat}_{d_n}, 
\ee
where $d=d_1\cdots d_n$.

A state $\rho$ is called {\it separable} (w.r.t.\ a given tensor product structure) if it can be written as 
$$
\rho=\sum_{i=1}^r \rho_i^{(1)}\otimes \cdots \otimes \rho_i^{(n)} 
\quad \mbox{with all }\rho_i^{(j)}\succcurlyeq 0. 
$$ 
This is obviously a stronger requirement than  $\rho$ being psd --- not every $\rho$ is separable.
Separable states are not too interesting from a  quantum information perspective: {\it not} separable states are called {\it entangled}, and entanglement is necessary for many quantum information tasks.

\medskip
\begin{center}
\deco
\end{center}
\medskip

A {\it quantum channel} is the most general transformation on quantum states. Mathematically,  it is modelled by a linear trace-preserving map 
$$ 
T\colon {\rm Mat}_d\to {\rm Mat}_s
$$ 
 that is  {\it completely positive}. Complete positivity means that the maps 
$$
{\rm id}_n\otimes T \colon   {\rm Mat}_{nd} 
\to 
{\rm Mat}_{ns}
$$ 
are positive (i.e.\ map psd matrices to psd matrices) for all $n$, where ${\rm id}_n$ is the identity map on  ${\rm Mat}_{n}$.

Any linear map $T\colon {\rm Mat}_d\to {\rm Mat}_s$ is uniquely determined by its {\it Choi matrix} 
 $$C_T:=\sum_{i,j=1}^d E_{ij}\otimes T\left(E_{ij}\right) \in {\rm Mat}_{ds},
 $$ 
 where $E_{ij}$ is the  matrix with a 1 in the $(i,j)$-position  and 0 elsewhere.\footnote{If this is expressed in the so-called computational basis (which is one specific orthonormal basis),  this is  written $E_{ij}=|i\rangle \langle j|$ in quantum information.}  
It is a basic fact that $T$ is completely positive if and only if $C_T$ is psd (see for example \cite{Pau02,Wo11}). 
Moreover, a completely positive map 
$T$ is \emph{entanglement-breaking}  \cite{Ho03} if and only if $C_T$ is a separable matrix, and 
$T$ is a positive map if and only if $C_T$ is \emph{block positive}, i.e. 
$$
\tr((\sigma\otimes \tau) C_T)\geqslant 0 \quad \textrm{for all } \sigma\succcurlyeq 0, \tau\succcurlyeq 0 .
$$
Note that this is weaker than $C_T$ being psd, in which case $\tr(\chi C_T)\geqslant 0$ for all $\chi\succcurlyeq0$ 
(see \cref{tab:pos}). 
We also remark that this link between positivity notions  of linear maps and their  Choi matrices does not involve the normalisation conditions on the maps (e.g.\ preserving the trace) or the matrices (e.g.\ having a given trace).

\begin{table}[ht]\centering
\begin{tabular}{|l|l|}\hline
\centering Linear map & Element in tensor product space \\
 $T : {\rm Mat}_d\to {\rm Mat}_s$ & 
 $\rho \in {\rm Mat}_d\otimes {\rm Mat}_s$\\
\hline \hline
Entanglement-breaking map & Separable matrix\\
Completely positive map &  Positive semidefinite matrix \\
Positive map & Block positive matrix\\ \hline
\end{tabular}
\caption{Correspondence between notions of positivity for linear maps and their Choi matrices. 
Entanglement-breaking maps are a subset of completely positive maps, which are a subset of positive maps. The same is true for the right column, of course. }
\label{tab:pos}
\end{table}

\subsection{Basic concepts from (free) semialgebraic geometry} 
\label{ssec:fsg}
We now introduce some basic concepts from free (i.e.\ noncommutative) semialgebraic geometry. For a slightly more detailed introduction, see \cite{Ne19} and references therein.

Our setup starts by considering a $\C$-vector space $V$ with an involution $*$. 
The two relevant examples are, first, the case where  $V$ is the space of matrices  
and * is the transposition with complex conjugation --- denoted $\dagger$ in quantum information ---, and, second,  
$\C^d$ with entrywise complex conjugation. 

The fixed points of the involution are called self-adjoint, or Hermitian, elements. We denote the set of Hermitian elements of $V$ by $V_{\rm her}$.
This is an $\R$-subspace of $V$, in which the {\it real} things happen.\footnote{Because it is where  positivity and other interesting phenomena happen.}

In the \emph{free} setup, we do not only consider $V$ but also higher levels thereof. Namely, for any $s\in\N$, we consider the space of $s\times s$-matrices with entries over $V$, 
$$
{\rm Mat}_s(V)= V\otimes {\rm Mat}_s. 
$$ 
Recall that  ${\rm Mat}_s$ refers to $s\times s$-matrices with entries over $\C$. 
${\rm Mat}_s(V)$  is a $\C$-vector space with a `natural' involution, consisting of transposing and applying $*$ entrywise. 
This thus promotes $V$ and $*$ to an entire hierarchy of levels, namely ${\rm Mat}_s(V)$ for all $s\in \N$ with the just described involution.

We are now ready to define the most general notion of a {\it  free real set}. This is nothing but  
a collection 
$$ 
\mathcal C=\left( C_s\right)_{s\in\N}
$$ 
where each $  C_s\subseteq {\rm Mat}_s(V)_{\rm her}=:{\rm Her}_s(V)$. We call $C_s$ the {\it set at level $s$}.

To make things more interesting, one often imposes conditions that connect the levels. 
One important example is {\it free convexity}, which is defined as follows. 
For any
$$\tau_i\in C_{t_i} \quad \mbox{with} \quad i=1,\ldots, n,$$ 
and 
\be
v_i\in{\rm Mat}_{t_i,s}\quad \mbox{with} \quad  \sum_{i=1}^n v_i^*v_i=I_s, 
\label{eq:norm}
\ee
it holds that
\be
\sum_{i=1}^n v_i^*\tau_iv_i\in C_s.
\label{eq:freeconvex}
\ee
Note that in \eqref{eq:freeconvex}  matrices over the complex numbers (namely $v_i$) are multiplied with matrices over $V$  (namely $\tau_i$). 
This is defined as matrix multiplication in the usual way for $v_i^*\tau_i v_i $, 
and using that elements of $V$ can be multiplied with complex numbers and added.
For example, for $n=1$, $t=2$, $s=1$,  $\tau= (\mu_{i,j})$ with $\mu_{i,j}\in V$ for $i=1,2$, and 
$v=(\lambda_1,\lambda_2)^t$ with $\lambda_i \in \C$, we have
$$
v^* \tau v = \sum_{i,j} \bar \lambda_i \lambda_j \mu_{i,j} . 
$$

Note that if free convexity holds, then every   $C_s$ is a convex set in the real vector space ${\rm Her}_s(V)$. 
But free convexity is generally a stronger condition than `classical' convexity,  as we will see.  

In addition, a conic version of free convexity is obtained when giving up the normalization condition on the $v_i$, i.e.\ the right hand side of Eq.\ \eqref{eq:norm}. In this case, $\mathcal C$ is called an {\it abstract operator system} (usually with the additional assumption that every $C_s$ is a proper convex cone).

\medskip
\begin{center}
\deco
\end{center}
\medskip

Now, {\it free semialgebraic sets} are free sets arising from polynomial inequalities. This will be particularly important for the connection we hope to illustrate in this paper.
In order to define these, take $V=\C^d$ with the involution provided by entrywise conjugation, so that  $V_{\rm her}=\R^d$. 
Let $z_1,\ldots, z_d$ denote free variables, that is, noncommuting variables. 
We can imagine each $z_i$ to represent a matrix of \emph{arbitrary size} --- later we will substitute  $z_i$ by a matrix of a given size, and this size will correspond to the level of the free semialgebraic set.

Now let  $\omega$ be a finite \emph{word} in the letters $z_1,\ldots, z_d$, that is, an ordered tuple of these letters. For example, $\omega$ could be $z_1 z_1 z_4$ or $z_4 z_5 z_4$. 
In addition, let $\sigma_\omega\in {\rm Mat}_m$ be a  matrix (of some fixed size $m$) that specifies the coefficients of word $\omega$; this is called the coefficient matrix. 
A {\it matrix polynomial} in the free variables $z_1,\ldots, z_d$ is an expression 
$$
p=\sum_{\omega} \sigma_\omega \otimes \omega, 
$$ 
where the sum is over all finite words $\omega$, and where only finitely many coefficient matrices $\sigma_\omega$ are nonzero. 

We denote the reverse of word $\omega$ by $\omega^*$. For example, if $\omega = z_1 z_2 z_3$ then $\omega^* = z_3 z_2 z_1$. 
In addition,  $(\sigma_\omega)^*$ is obtained by transposition and complex conjugation of $\sigma_\omega.$ 
If the coefficient matrices fulfill 
\be \label{eq:coeffmat}
(\sigma_\omega)^*=\sigma_{\omega^*}, 
\ee   
then for any tuple of Hermitian matrices 
$(\tau_1,\ldots, \tau_d)\in{\rm Her}_s^d$   
we have that 
$$
p(\tau_1,\ldots, \tau_d)=\sum_\omega\sigma_\omega\otimes \omega(\tau_1,\ldots, \tau_d)\in {\rm Her}_{ms}. 
$$ 
That is, $p$ evaluated at the Hermitian matrices $\tau_1,\ldots, \tau_d$ is a Hermitian matrix itself. 

So, for a given matrix polynomial $p$ satisfying condition \eqref{eq:coeffmat}, 
we define the free semialgebraic set at level $s$ 
as the set of Hermitian matrices of size $s$ 
such that $p$ evaluated at them is psd:  
$$
C_s(p):=\left\{ (\tau_1,\ldots, \tau_d)\in {\rm Her}_s^d\mid p(\tau_1,\ldots, \tau_d)\succcurlyeq 0\right\} .
$$
Finally we define the free semialgebraic set as the collection of all such levels:
 $$
 \mathcal C(p):=\left(C_s(p)\right)_{s\in\N}.
 $$ 
 
For example, let $E_{ii}$ denote the matrix with a 1 in entry $(i,i)$ and 0 elsewhere.
Then the matrix polynomial 
$$
p=\sum_{i=1}^d E_{ii}\otimes z_i
$$ 
defines the following free semialgebraic set at level $s$  
$$
C_s(p) =
\left\{ (\tau_1,\ldots,\tau_d)\in {\rm Her}_s^d\mid E_{11} \otimes \tau_1 + \ldots  + E_{dd} \otimes \tau_d\succcurlyeq 0\right\} .
$$ 
The positivity condition is equivalent to $\tau_i\succcurlyeq 0$ for all $i$, which gives this free set the name  {\it free positive orthant}.
Note that for $s=1$, the `free' variables become  real numbers,  
 $$
C_1(p) =
\left\{ (a_1,\ldots, a_d)\in \R^d \mid a_i \geqslant 0  \:\:\forall i\right\} , 
$$
which defines the positive orthant in $d$ dimensions. 

It is easy to see that any free semialgebraic set is closed under direct sums, meaning that if 
$(\tau_1,\ldots, \tau_d)\in C_s(p)$, 
$(\chi_1,\ldots, \chi_d)\in C_r(p)$ 
then 
$$(\tau_1 \oplus \chi_1,\ldots ,\tau_d\oplus \chi_d) \in C_{r+s}(p),$$ where $\tau_i \oplus \chi_i$ denotes
the block diagonal sum of two Hermitian matrices. 
This is because 
$ 
p(\tau_1 \oplus \chi_1,\ldots \tau_d\oplus \chi_d) = p(\tau_1,\ldots \tau_d) \oplus p(\chi_1,\ldots \chi_d), 
$  
which is psd if and only if each of the terms is psd. 

Note also that a {\it semialgebraic set} is a Boolean combination of $C_1(p_i)$ for  a finite set of polynomials $p_i$. A `free semialgebraic set' is thus a noncommutative generalisation thereof, with the difference that usually  a single polynomial $p$ is considered.

\medskip
\begin{center}
\deco
\end{center}
\medskip

\begin{figure}[t]
  \centering
 \begin{minipage}[]{0.2\textwidth}
    \includegraphics[width=\textwidth]{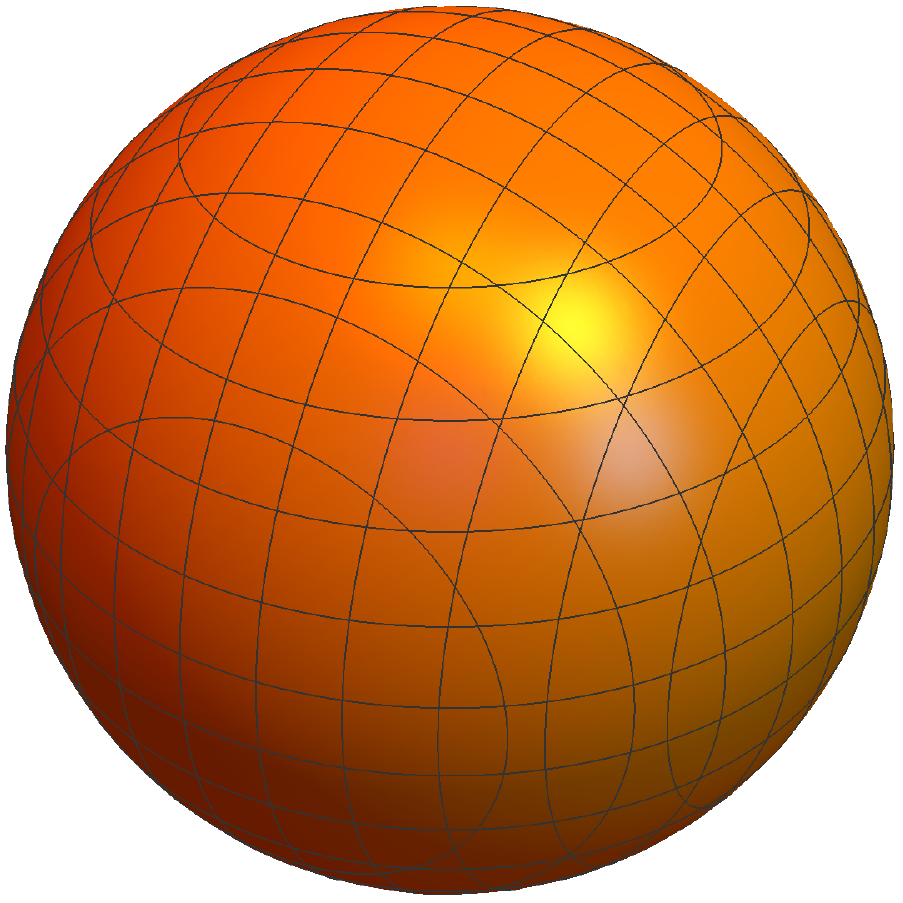}
      \end{minipage}
 \hspace{0.1\textwidth}
  \begin{minipage}[]{0.2\textwidth}
    \includegraphics[width=\textwidth]{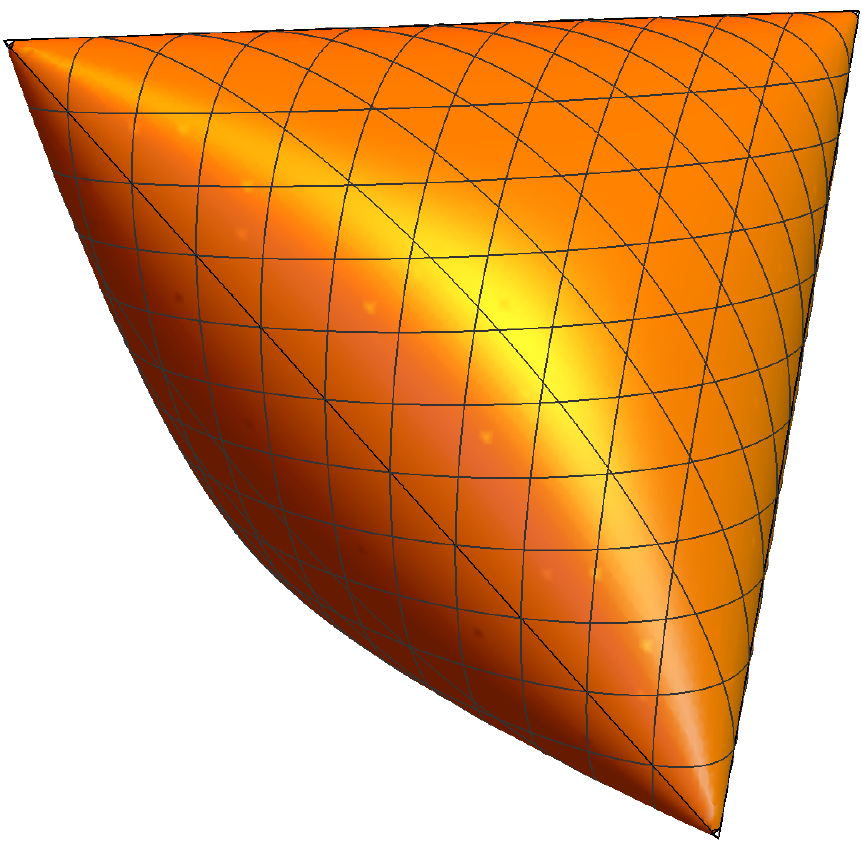}
    \end{minipage}
     \hspace{0.1\textwidth}
 \begin{minipage}[]{0.2\textwidth}
  \includegraphics[width=1.2\textwidth]{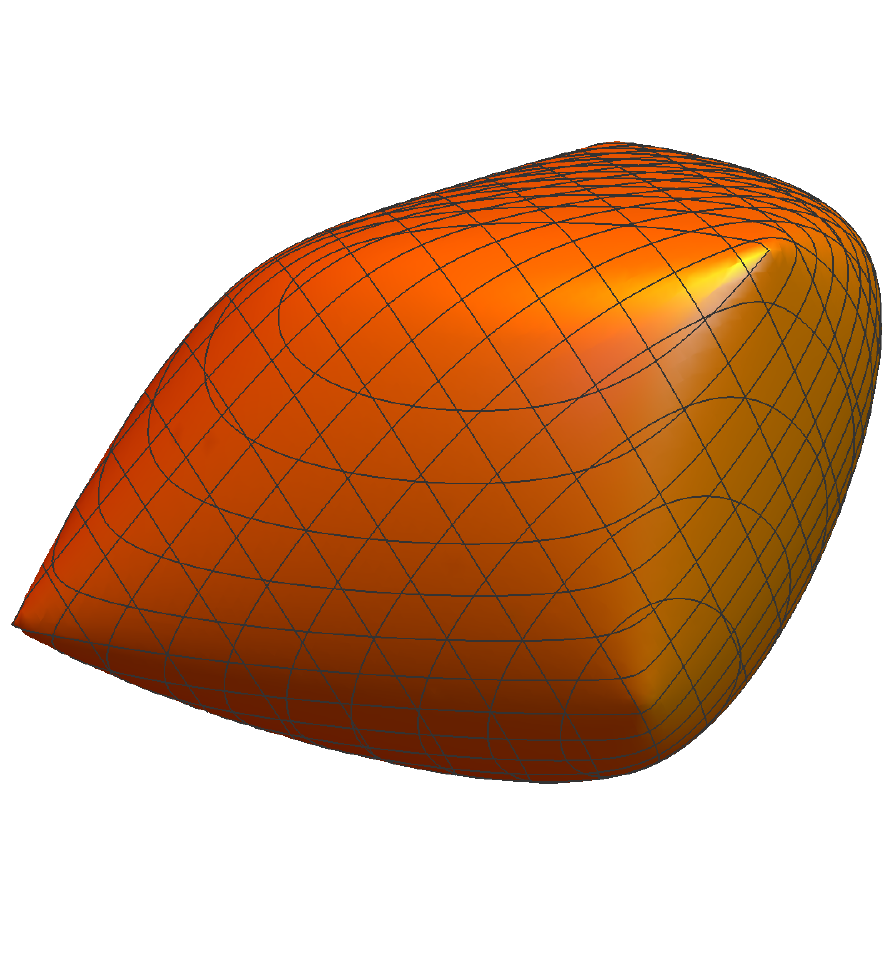}
    \end{minipage}
\caption{Some three-dimensional spectrahedra taken from \cite{NP}. 
Spectrahedra are convex sets described by a linear matrix inequality, and polyhedra are particular cases of  spectrahedra. }
\label{fig:spec}
\end{figure}

A very special case of free semialgebraic sets are \emph{free spectrahedra}, which arise from linear matrix polynomials. A {\it linear matrix polynomial} is a matrix polynomial where every word $\omega$ depends  only on one variable, i.e. 
$$
\ell=\sigma_0\otimes 1+\sum_{i=1}^d \sigma_i\otimes z_i
$$ 
with 1 being the empty word, and all $\sigma_i\in {\rm Her}_m$. 
The corresponding free set at level $s$ is given by 
$$
C_s(\ell)=\left\{ (\tau_1,\ldots, \tau_d)\in {\rm Her}_s^d\mid \sigma_0\otimes I_s + \sum_{i=1}^d \sigma_i\otimes \tau_i\succcurlyeq 0\right\}
$$ 
and $\mathcal C(\ell)$ is called a {\it free spectrahedron}. The first level set, 
$$
C_1(\ell)=\left\{ (a_1,\ldots, a_d)\in \R^d\mid \sigma_0+ a_1\sigma_1+\cdots +a_d\sigma_d\succcurlyeq 0\right\},
$$ 
is known as a {\it classical spectrahedron}, or simply, a  {\it  spectrahedron} 
 (see  \cref{fig:spec} for some three-dimensional  spectrahedra). 
If all $\sigma_i$ are diagonal in the same basis, then the  spectrahedron $C_1(\ell)$ becomes a polyhedron. (Intuitively, polyhedra have flat facets whereas the borders of spectrahedra can be round, as in \cref{fig:spec}.)
Thus, every polyhedron is a spectrahedron, but not vice versa. 

While the linear image (i.e.\ the \emph{shadow}) of a polyhedron is a polyhedron, the shadow of a spectahedron need not be a spectahedron. 
The forthcoming book \cite{NP} presents a comprehensive treatment of spectrahedra and their shadows.\footnote{Shadows can be very different from the actual thing, as \href{https://www.arch2o.com/wp-content/uploads/2012/09/Arch2O-light-and-shadow-kumi-yamashita-3.jpg}{this shadow art} by Kumi Yamashita shows.}

\section{One wonderland through two looking glasses}
\label{sec:results} 

Let us now explain some recent results that illustrate how concepts and methods from the two disciplines interact.  
We will focus on positivity and separability  (\cref{ssec:pos}), 
quantum magic squares (\cref{ssec:magic}), 
non-local games (\cref{ssec:game}), 
and positivity in tensor networks (\cref{ssec:tn}). 

\subsection{Positivity and separability}
\label{ssec:pos}

For fixed  $d,s\in\N$ consider the set of  states and separable states in ${\rm Mat}_d\otimes {\rm Mat}_s$, namely ${\rm State}_{d,s}$ and ${\rm Sep}_{d,s}$, 
 respectively. 
Both sets are closed  in the real vector space 
$
{\rm Her}_d\otimes {\rm Her}_s
$. Moreover, both are semialgebraic, since 
  $ {\rm State}_{d,s}$ is a classical spectrahedron, and   ${\rm Sep}_{d,s}$ can be proven to be semialgebraic using the {\it projection theorem/quantifier elimination} in the theory of real closed fields (see, e.g., \cite{PD01}).

It has long been known that ${\rm Sep}_{d,s}$ is a strict subset of ${\rm State}_{d,s}$  whenever $d,s >1 $. 
A recent work  by Fawzi \cite{Faw19}, building on Scheiderer's  \cite{Sc18}, 
strengthens this result, by showing that the geometry of these two sets is significantly different:

\begin{theorem}[\cite{Faw19}] \label{thm:sepshadow}
If $d+s>5$ then ${\rm Sep}_{d,s}$ 
 is not a spectrahedral shadow. 
\end{theorem}

Recall that a spectahedral shadow is the linear image of a spectrahedron. 

Together with the relations of \cref{tab:pos}, 
it follows from the previous result 
that the corresponding sets of linear maps $T: {\rm Mat}_d \to {\rm Mat}_s$ satisfy that: 
\begin{itemize}
\item[(i)] Entanglement-breaking maps form a convex semialgebraic set which is not a spectrahedral shadow, 
\item[(ii)]  Completely positive maps form a  spectrahedron, and 
\item[(iii)] Positive maps form a convex semialgebraic set  which is not a spectrahedral shadow. This follows from (i), the duality of positive maps and entanglement-breaking   maps, and the fact that duals of  spectrahedral  shadows are also spectrahedral shadows \cite{NP}. 
\end{itemize}

Let us now consider the set of states and separable states as free sets. Namely, for fixed $d\geqslant 1$ 
let 
$$
{\rm State}_d:=\left( {\rm State}_{d,s}\right)_{s\in\N} \quad\mbox{ and } \quad {\rm Sep}_{d}:=\left({\rm Sep}_{d,s}\right)_{s\in\N}.
$$ 
This is a particular case of the setup described above, where $V={\rm Mat}_d$ and the involution is provided by $\dagger$. 
Moreover, both sets satisfy the condition of free convexity (Eq.\ \eqref{eq:freeconvex}).  
In addition, ${\rm State}_d$ is a free spectrahedron, whereas ${\rm Sep}_{d}$ is not, since for fixed $s$ it is not even a classical  
spectrahedral shadow at level $s$ due to \cref{thm:sepshadow}.  

Viewing states as free sets also leads to an easy  conceptual proof  of the following result \cite{D19}, which was first proven by Cariello \cite{Car}.  

\begin{theorem}[\cite{D19,Car}]
For arbitrary $d,s\in\N$, if $\rho\in {\rm State}_{d,s}$ is of tensor rank $2$, i.e.\ it can be written as 
\be
\rho=\sigma_1\otimes \tau_1 + \sigma_2\otimes \tau_2, 
\label{eq:rho}
\ee
where $\sigma_i$ and $\tau_i$ are Hermitian, 
then it is separable. 
\end{theorem}

Note that $\sigma_i$ and $\tau_i$ need not be psd.
Let us sketch the proof of \cite{D19} to illustrate the method.

\begin{proof}
Consider the linear matrix polynomial 
$\ell=\sigma_1\otimes z_1+ \sigma_2\otimes z_2$, where $\sigma_1,\sigma_2$ are given in Eq.\ \eqref{eq:rho}. 
The fact that $\rho$ is a state  means that the corresponding free set of level $s$ contains $(\tau_1,\tau_2)$: 
$$
(\tau_1,\tau_2)\in C_s(\ell).
$$ 

At level one, the spectrahedron $C_1(\ell)$ is a convex cone in $\R^2$. 
A convex cone in the plane must be a \emph{simplex cone}, i.e.\  a cone whose number of extreme rays equals the dimension of the space.  In $\R^2$
this means that the cone is spanned by two vectors,  
$$
C_1(\ell)={\rm cone}\{v_1,v_2\}, 
$$ 
where $v_1, v_2\in\R^2$. 
When the cone at level one is a simplex cone,  the  free convex cone is fully determined \cite{ev,FNT}.

In addition,  the sets
$$
T_s:=\left\{ v_1\otimes \eta_1+v_2\otimes\eta_2\mid 0\preccurlyeq \eta_i\in{\rm Her}_s\right\}
$$ 
also give rise to a free convex cone $\left( T_s\right)_{s\in\N}$, and we have that $T_1 = C_1(\ell)$. 

These two facts imply that  $T_s=C_s(\ell)$ for all $s\in\N$. 
Using a representation for $(\tau_1,\tau_2)$ in $T_s$, 
and substituting into Eq.\ \eqref{eq:rho} results in a separable decomposition of $\rho$. 
\end{proof}

The crucial point in the proof is that when the cone at level one is a simplex cone,  the  free convex cone is fully determined. 
This is not a very deep insight --- it can easily be reduced to the case of the positive orthant, where it is obvious. 

Note that the separable decomposition of $\rho$ obtained in the above proof contains only   two terms --- in the language of \cite{D19,DN}, $\rho$ has separable rank 2.

\bigskip 
\begin{center}
 \deco 
 \end{center}
\bigskip 

References \cite{BN1,BN2} also propose to use free spectrahedra to study some problems in quantum information theory, but from a different perspective. 
Given $d$ Hermitian matrices $\sigma_1,\ldots, \sigma_d\in{\rm Her}_m$, one would like to know  whether they fulfill 
$$
0 \preccurlyeq \sigma_i \preccurlyeq  I_m,
$$
 because this implies that  each $\sigma_i$ gives rise to  the binary POVM consisting of $\sigma_i,I_m-\sigma_i$.  
 In addition, one would  like to know  whether $\sigma_1,\ldots, \sigma_d$ are {\it jointly measurable}, meaning that these  POVMs are the marginals   of one   POVM (see \cite{BN1} for an exact definition).

Now use $\sigma_1,\ldots, \sigma_d$ to construct the linear matrix polynomial 
$$
\ell:=I_m\otimes 1-\sum_{i=1}^d (2\sigma_i-I_m)\otimes z_i
$$ 
and consider its  free spectrahedron $\mathcal C(\ell)=\left(C_s(\ell)\right)_{s\in\N}.$
Define the {\it matrix diamond} as the free spectrahedron $\mathcal D=\left(D_s\right)_{s\in\N}$ with 
$$
D_s:=\left\{ (\tau_1,\ldots, \tau_d)\in {\rm Her}_s^d\mid I_s-\sum_{i=1}^d \pm \tau_i \succcurlyeq 0\right\},
$$
 where all possible choices of signs $\pm$ are taken into account. 
 Note that $D_1$ is just the unit ball of $\R^d$ in $1$-norm, which explains the name {\it diamond}.  
Note also that  $D_1\subseteq S_1(\ell)$ is equivalent to  $0 \preccurlyeq \sigma_i \preccurlyeq  I_m$ for all $i=1,\ldots, d$. 
Since these finitely many conditions can be combined  into a single linear matrix inequality (using diagonal blocks of matrix polynomials), $\mathcal D$ is indeed a free spectrahedron. 
 The following result translates the joint measurability to the containment of free spectrahedra:

\begin{theorem}[\cite{BN1}]
$\sigma_1,\ldots, \sigma_d$ are jointly measurable if and only if 
 $\mathcal D\subseteq\mathcal C(\ell)$. 
\end{theorem}

That one free spectrahedron is contained in another, $\mathcal D\subseteq\mathcal C(\ell)$, means that each of their corresponding levels satisfy the same containment, i.e.\  $D_s\subseteq C_s(\ell)$ for all $s\in\N$. 

The containment of spectrahedra and free spectrahedra has received considerable attention recently \cite{bental, hecp,  hedi, FNT, PSS}. One often studies {\it inclusion constants} for containment, 
which determine how much the small spectrahedron needs to be shrunk  in order to obtain inclusion. In \cite{BN1,BN2} this is used to quantify the degree of incompatibility, and to obtain lower bounds on the joint measurability  of quantum measurements.

\subsection{Quantum magic squares} 
\label{ssec:magic}
Let us now look at magic squares and their quantum cousins. 

\begin{figure}[t]
\centering
 \begin{minipage}[]{0.4\textwidth}
\includegraphics[width=\textwidth]{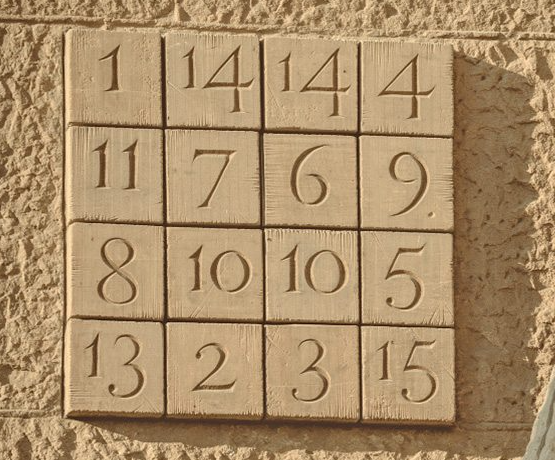}
      \end{minipage}  
      \hspace{0.15\textwidth}
       \begin{minipage}[]{0.33\textwidth}
\includegraphics[width=\textwidth]{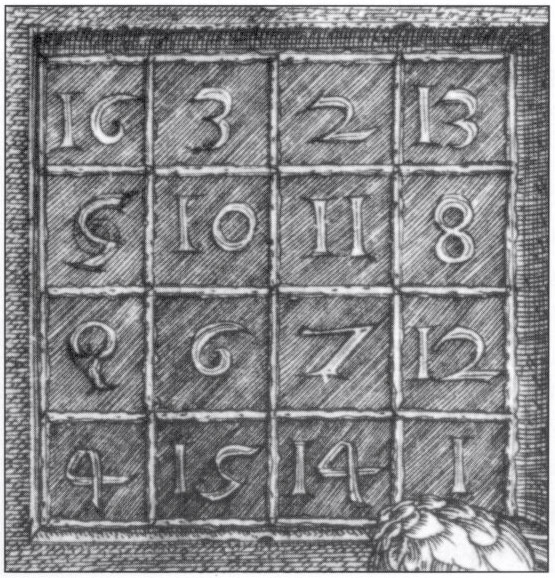}
      \end{minipage}
\caption{(Left) The magic square on the fa\c{c}ade of the Sagrada Fam\'ilia in Barcelona, 
where every row and column adds to 33. (Right) The magic square in Albrecht D\"urer's lithograph \emph{Melencolia I}, where  every row and column adds to 34. }
\label{fig:magicsquare}
\end{figure}

A {\it magic square} is a  $d\times d$-matrix with positive entries such that every row and column sums to the same number (see \cref{fig:magicsquare} for two beautiful examples.) 
A {\it doubly stochastic matrix} is a $d\times d$-matrix with real nonnegative entries, in which each row and each column sums to $1$. So doubly stochastic matrices contain a probability measure in each row and each column. 
 For example, dividing every entry of D\"urer's magic square by 34 results in a doubly stochastic matrix. 
Now, the set of doubly stochastic matrices forms a polytope, whose vertices consist of the {\it permutation matrices}, i.e.\ doubly stochastic matrices with a single 1 in every row and column and 0 elsewhere (that is, permutations of the identity matrix). This is the content of the famous Birkhoff--von Neumann Theorem. 

A  `quantum' generalization of a doubly stochastic matrix is obtained by putting a POVM (defined in \cref{ssec:qi}) in each row and each column of a $d\times d$-matrix. 
This defines a  {\it quantum magic square} \cite{D20}. 
That is, in passing from  doubly stochastic matrices to quantum magic squares, 
we promote the nonnegative numbers to  psd matrices. 
The normalisation conditions on the numbers (that they sum to 1) become the normalisations of the POVM (that they sum to the identity matrix).

What is a quantum generalisation of a permutation matrix? Permutation matrices only contain  0s and 1s, so in passing to the quantum version, we promote  0 and 1 to orthogonal projectors (given  that 0 and 1 are the only numbers that square to themselves). The relevant notion is thus that of a \emph{projection valued measure} (PVM), 
in which each measurement operator $\tau_1,\ldots, \tau_d$ is an orthogonal projection,  
 $\tau_i^2=\tau_i$.  
{\it Quantum permutation matrices}  are magic squares containing a PVM in each row and column \cite{Ba}.\footnote{
See the closely related notion of \emph{quantum Latin squares} \cite{Mu16,Ho20}, which in essence are quantum permutation matrices with rank 1 projectors.}

While PVMs are a special case of POVMs,  every POVM {\it dilates} to a PVM
(see, e.g., \cite{Pau02}):

\begin{theorem}[Naimark's Dilation Theorem]
Let $\tau_1,\ldots,\tau_d$ (of size $m\times m$) 
 form a POVM. 
Then there exists a PVM $\sigma_1,\ldots, \sigma_d$ (of size $n\times n$, for some $n$)
and a matrix $v\in {\rm Mat}_{n,m}$ such that 
$$
v^*\sigma_iv=\tau_i \mbox{ for all }i=1,\ldots, d.
$$
\end{theorem}
In terms of free sets, this theorem states that \emph{the free convex hull of the set of PVMs is precisely the set of POVMs}. Both sets are free semialgebraic, and the POVMs even form a free spectrahedron.

Through the glass of free semialgebraic geometry, quantum magic squares form a free spectrahedron over the space $V={\rm Mat}_d$, equipped with entrywise complex conjugation as an involution. 
Level $s$ corresponds to  POVMs with matrices of size $s\times s$, and thus level 1  corresponds to doubly stochastic matrices. 
We thus recover the magic in the classical world at level 1, 
and we have an infinite tower of levels on top of that  expressing the quantum case.

Furthermore, quantum permutation matrices form a free semialgebraic set whose first level consists of permutation matrices. 
The  `classical magic' is thus again found at level 1, and the quantum magic is expressed in an infinite tower on top of it.

Now, recall that the Birkoff--von Neumann theorem says that the convex hull of the set of  permutation matrices is the set of doubly stochastic matrices. 
So the permutation matrices are the vertices of the polytope of doubly stochastic matrices. 
In the light of the towers of quantum magic squares and quantum permutation matrices, this theorem fully characterises what happens at level one. 
We ask whether a similar characterisation is possible for the quantum levels:  
\emph{Is the free convex hull of quantum permutation matrices equal to the set of quantum magic squares?}  

This question can be phrased in terms of dilations as follows. By Naimark's Dilation Theorem  we know that every POVM dilates to a PVM. The question is whether this also holds for a two-dimensional array of POVMs, i.e.\ whether every square of POVMs can dilated to a square of PVMs. 
The non-trivial part is that the dilation must work simultaneously for all POVMs in the rows and columns. 
The two-dimensional version of Naimark's Dilation Theorem can thus be phrased as: 
\emph{Does every quantum magic square dilate to a quantum permutation matrix?} 

The answer to these questions is `no': 
these quantum generalisations fail to be true  in the simplest nontrivial case. This means that  there must exist very strange (and thus very interesting) quantum magic squares:

\begin{theorem}[\cite{D20}] For each $d\geqslant 3$, the free convex hull of the free semialgebraic set of $d\times d$ quantum permutation matrices is strictly contained in the free spectrahedron of quantum magic squares. This strict containment already appears at level $s=2$. 
\end{theorem}

The latter statement means that there is a $d\times d$-matrix with POVMs of size $2\times 2$ in each row and column which does not dilate to a matrix with a PVM in each row and column.

In words, the tower of quantum levels does not admit the same kind of `easy' characterisation as level one or the case of a single POVM --- at least not the natural generalisations we have considered here. 
This is yet another sign of the richer structure of the quantum world compared to the classical one.

\subsection{Non-local games and quantum correlations} 
\label{ssec:game}
Consider a game with two players, Alice and Bob, and a referee.
The referee chooses a question randomly from finite sets $\mathcal Q_A$ and $\mathcal Q_B$ for Alice and Bob, respectively, 
and sends them to Alice and Bob. 
Upon receiving her question, Alice  chooses from a finite set $\mathcal A_A$ of answers, 
and similarly Bob chooses his answer from the finite set $\mathcal A_B$. 
They send their answers to the referee, who computes a winning function 
$$
w\colon \mathcal Q_A\times \mathcal Q_B\times \mathcal A_A\times \mathcal A_B\to \{0,1\}
$$ 
to determine whether they win or lose the game (value of $w$ being 1 or 0, respectively).

During the game, Alice and Bob know both the winning function  $w$ and the probability measure on $\mathcal Q_A\times \mathcal Q_B$ used by the referee to choose the questions. 
So before the game starts Alice and Bob agree on a joint strategy. 
However, during the game Alice and Bob are `in separate rooms' (or in separate galaxies) so they cannot communicate. 
In particular, Alice will not know Bob's question and vice versa.
In order to find the strategy that maximises the winning probability, Alice and Bob  have to solve an optimisation problem.\footnote{Thus, strictly speaking,  this is not a game in the game-theoretic sense, but (just) an optimisation problem.}

What kind of strategies may Alice and Bob  choose? It depends on the resources they have. 
First, in a {\it classical deterministic strategy}, both Alice and Bob  reply deterministically to each of their questions, and they do so independently of each other. 
This is described by two functions 
$$
c_A\colon\mathcal Q_A\to \mathcal A_A  \quad \mbox{ and } \quad c_B\colon \mathcal Q_B\to \mathcal A_B,
$$ 
which specify which answer Alice and Bob give to each question. 

Slightly more generally, in a {\it classical randomised strategy}, Alice and Bob's answers are probabilistic, but still independent of each other. This is described by 
 $$
 r_A\colon\mathcal Q_A\to \mathcal {\rm Pr}(\mathcal A_A)  \quad \mbox{ and } \quad r_B\colon \mathcal Q_B\to \mathcal {\rm Pr}(\mathcal A_B),
 $$ 
 where ${\rm Pr}(S)$ denotes the set of probability measures on the set $S$. 
 Namely,  if Alice receives question $a$, the probability that she answers $x$ is given by $r_A(a)(x)$, where  $r_A(a)$ is the probability measure on $\mc{A}_A$ corresponding to question $a$. 
 Similarly, Bob answers $y$ to $b$ with probability $r_B(b)(y)$. 
Since Alice and Bob answer independently of each other, the joint probability of answering $x,y$ upon questions $a,b$ is the product of the two, 
\be\label{eq:crs}
p(x,y\mid a,b)=r_A(a)(x)\cdot r_B(b)(y). 
\ee

Finally, a {\it quantum strategy} allows them to share a bipartite state  $\rho\in {\rm State}_{d,s}$.  
The questions determine which measurement to apply to their part of the state, and the measurement outcomes determine the answers.  
This is described by  functions 
\be\label{eq:quantstrat}
q_A\colon\mathcal Q_A\to {\rm POVM}_d(\mathcal A_A)\quad \mbox{ and } \quad q_B\colon \mathcal Q_B\to {\rm POVM}_s(\mathcal A_B)
\ee
whose image is the set of POVMs with matrices of size $d\times d$ and $s\times s$, respectively, on the respective sets of answers. 
The probability that Alice answers $x$ upon receiving $a$ is described by $q_A(a)(x)$, which is  the psd matrix that the POVM $q_A(a)$ assigns to answer $x$. 
Similarly, Bob's behaviour is modelled by $q_B(b)(y)$. 
Since they act independently of each other, this is described by the tensor product of the two. Using rule \eqref{eq:povm}, we obtain that their joint probability is given by 
\be\label{eq:qs}
p(x,y\mid a,b)=\tr\left(\rho  \left(q_A(a)(x)\otimes q_B(b)(y)\right)\right). 
\ee

Now, the table of conditional probabilities 
$$
\left(p(x,y\mid a,b)\right)_{(a,b,x,y)\in \mathcal Q_A\times \mathcal Q_B\times \mathcal A_A\times \mathcal A_B}
$$ 
is called the {\it correlation matrix } of the respective strategy. 
For any given kind of strategy, 
the set of correlation matrices is the feasible set of the optimisation problem that Alice and Bob have to solve. 
The objective function of this optimisation problem is given by the winning probability. 
Since this objective function is linear in the correlation matrix entries,  
one can replace the feasible set by its convex hull.  

The important fact is that quantum strategies cannot be reproduced by classical randomised strategies:  

\begin{theorem}[\cite{Be,Cl69}] \label{thm:bell}
If at least $2$ questions and $2$ answers exist for both  Alice and Bob, the convex hull of correlation matrices of classical randomised strategies is strictly contained in the set of correlation matrices of quantum strategies.
\end{theorem}

For   classical randomised strategies, passing to the convex hull has the physical interpretation of including a \emph{hidden variable}. The latter is a variable whose value is unknown to us, who are describing the system, and it is usually denoted $\lambda$. However, this mysterious variable $\lambda$ is shared between Alice and Bob, and it will determine the choice of their POVMs together with their respective questions $a,b$. This is the physical interpretation of the convex hull 
$$
p(x,y\mid a,b)= \sum_\lambda  q_\lambda \: r_A(a,\lambda)(x)\cdot r_B(b,\lambda)(y), 
$$
where $q_\lambda$ is the probability of the hidden variable taking the value $\lambda$. 
For example, we can imagine that Alice and Bob are listening to a radio station which plays songs from  a certain list, 
but this is a `private' radio station to which we have no access. 
The song at the moment of playing the game (i.e.\ receiving the questions) will determine the value of $\lambda$ (i.e.\ $\lambda$ is an index of that list).

\cref{thm:bell} thus states that  quantum strategies cannot be emulated by classical strategies, even if we take into account `mysterious' hidden variables.

\medskip
\begin{center}
\deco
\end{center}
\medskip 

Let us now approach these results from the perspective of free sets. 
Assume for simplicity that all four  sets $\mathcal Q_A,\mathcal Q_B,\mathcal A_A,\mathcal A_B$  have two elements. 
A quantum strategy consists of a state $\rho\in {\rm State}_{d,s}$ and the following psd matrices for Alice and Bob, respectively,  satisfying this normalisation condition:
\be \label{eq:normsigmatau}
\begin{split}
&\sigma_j^{(i)} \succcurlyeq 0  \mbox{ and } 
\tau_j^{(i)} \succcurlyeq 0 \\
&\mbox{such that } \sum_j \sigma_j^{(i)} =I_d\mbox{ and } 
\sum_j \tau_j^{(i)} =I_s,  
\end{split}
\ee  
where $i,j=1,2$.
The superscript refers to the questions and the subscript to the answers. 
The  correlation matrix is given by 
$$
\left(\tr\left(\rho \: \left(\sigma_k^{(i)}\otimes \tau_l^{(j)}\right)\right)\right)_{i,j,k,l} .
$$ 
Using the spectral decomposition  of  $\rho =\sum_r v_r v_r^*$, 
it can be written as 
\be \label{eq:fch1}
\left(\sum_r v_r^* \left(\sigma_k^{(i)}\otimes \tau_l^{(j)}\right)v_r\right)_{i,j,k,l}, 
\ee
where  $v_r\in \C^d\otimes \C^s$ and $\sum_r v_r^*v_r=1$. 
Through the  looking glass  of free semialgebraic geometry,  
this is first level of  a free convex hull. 
To see this, define the free set $\mathcal I$ as 
\begin{align}
\mathcal{I} = 
\bigcup_{d,s\geq 1}
\left\{\left(\sigma_k^{(i)}\otimes \tau_l^{(j)}\right)_{i,j,k,l}\in {\rm Mat}_4({\rm Mat}_{d}\otimes {\rm Mat}_{s} ) 
\mid \sigma_k^{(i)} \mbox{and } \tau_l^{(j)} \mbox{satisfy }\eqref{eq:normsigmatau}
\right\}
\end{align}
(Note that the 4 is due the fact that we have 2 questions and 2 answers; more generally we would have a matrix of size 
$|\mathcal{Q}_A||\mathcal{Q}_B| \times |\mathcal{A}_A||\mathcal{A}_B|$. Note also that the ordering of questions and answers of Alice and Bob is irrelevant for the following discussion.)

If we look at level 1 of this free set, we encounter that $I_1$  is the subset of ${\rm Mat}_4(\R)$  consisting precisely of the correlation matrices of classical randomized strategies. In other words, when $d=s=1$, the formula coincides with that of  \eqref{eq:crs}. 
Furthermore, higher levels of this free set contain the tensor products of POVMs of Alice and Bob in the corresponding space ${\rm Mat}_{d}$ and ${\rm Mat}_{s}.$ 
Since $I_1$ is called the {\it independence model} in algebraic statistics \cite{Dr}, we call  $\mathcal I$ the {\it free independence model}, since this is the natural noncommutative generalisation of independent strategies.

Let us now consider the free convex hull of $\mathcal{I} $. 
First of all, computing the conditional probabilities of a  pair of POVMs with a given state $\rho$ corresponds 
to compressing to level 1 with the vectors $\{v_r\}$ given by the spectral decomposition of $\rho$, as in \eqref{eq:fch1}.
So \emph{the set of quantum correlations is the first level of the free convex hull of the free independence model.}

We thus encounter an interesting phenomenon: 
the free convex hull of a free set can be larger than the classical convex hull at a fixed level. 
Specifically, the convex hull of $I_1$ is the set of classical correlations, 
whereas the free convex hull of $\mathcal{I}$ at level 1 is the set of quantum correlations, 
which are different by \cref{thm:bell}. 
In fact, wilder things can happen: fractal sets can arise in the free convex hull of free semialgebraic sets  \cite{al}. 
We wonder what these results imply for the corresponding quantum information setup.

Now, in the free convex hull of $\mathcal{I}$, what do higher levels  correspond to?
Compressing to lower levels (i.e.\ with smaller  $ds$)  corresponds to taking the partial trace with a psd matrix of size smaller than $ds$. This results in 4 psd matrices (one for each $i,j,k,l$), each  of size $<ds$, and which not need be an elementary tensor product. 

What about `compressing' to higher levels?
Any compression to a higher level can be achieved by   direct sums of the POVMs of Alice and Bob and a compression to a lower level as we just described. 
The number of elements in this direct sum is precisely $n$ in \eqref{eq:freeconvex}.
Another way of seeing that the direct sum is needed is by noting that, if $n=1$, the matrices $v_i$ cannot fulfill the normalisation condition on the right hand side of \eqref{eq:freeconvex}. 
In quantum information terms, this says that a POVM in a given dimension cannot be transformed to a POVM in a larger dimension by means of an isometry, because the terms will sum to a projector instead of the identity.

Let us make two final remarks. 
The first one is that $\mathcal I$ is not a free semialgebraic set, for the simple reason that it is not closed under direct sums (which is a property of these sets, as we saw in \cref{ssec:fsg}), 
as is easily checked. 

The second remark is that the free convex hull of the free independence model is not closed. 
This follows from the fact that, at level 1, this free convex hull fails to be closed, as shown in  \cite{sl}  and for smaller sizes in \cite{dy}. 

\begin{theorem}[\cite{sl,dy}]
For at least $5$ questions and $2$ answers, the set of quantum correlation matrices is not closed.
\end{theorem}

In our language, this implies that the level $ds$ --- which is to be compressed to level 1 in the construction of the free convex hull --- cannot be upper bounded. That is, the higher $ds$ the more things we will obtain in its compression to level 1. 

In the recent preprint \cite{Fu21} the membership problem in the closure of the set of quantum correlations is shown to be undecidable, for a fixed (and large enough) size of the sets of questions and answers.

\medskip
\begin{center}\deco\end{center}
\medskip

A computational approach to quantum correlations, comparable  to sums-of-squares and moment relaxation approaches in polynomial optimisation \cite{blek,NP},  is the {\it NPA-hierarchy} \cite{NPA1,NPA2,NPA3}. We briefly describe the approach here, omitting technical details. Assume one is given a table 
$$
p=\left(p(x,y\mid a,b)\right)_{(a,b,x,y)\in \mathcal Q_A\times \mathcal Q_B\times \mathcal A_A\times \mathcal A_B}, 
$$
 and the task is to check whether it is the correlation matrix of a quantum strategy.
The NPA hierarchy provides a family of necessary conditions, each more stringent than the previous one, for $p$ to be a quantum strategy.

In order to understand the NPA hierarchy, we will first \emph{assume} that $p$ is a correlation matrix, i.e.\ there is a state $\rho$ and  strategies such that \eqref{eq:qs} holds. 
We will use this state and strategies to define a positive functional on a certain algebra. 
Namely, we consider the {\it game $*$-algebra} 
$$
\mathcal G:=\C\langle \mathcal Q_A\times \mathcal A_A,\mathcal Q_B\times \mathcal A_B\rangle. 
$$ 
This is an  algebra of polynomials in certain noncommuting variables.  
Explicitly,  for each question and answer pair from Alice and Bob, 
$(a,x)$ 
and $(b,y)$, 
there is an associated self-adjoint variable, $z_{(a,x)}$ and  $z_{(b,y)}$, respectively.    
$\mathcal G$ consists of all polynomials with complex coefficients in these variables; for example, the monomial $z_{(a,x)} z_{(b,y)}\in \mathcal{G}$. 
Now, \emph{if we had the strategy $\rho,q_A, q_B$} we could construct a linear functional 
$$
\varphi\colon\mathcal G \to \C
$$  
by evaluating the variables $z_{(a,x)}$ and  $z_{(b,y)}$ at the psd  matrices $q_A(a)(x)\otimes I_s$ and 
$I_d\otimes q_B(b)(y)  $, respectively, 
and computing the trace inner product  with the state $\rho$. 
So, in particular, evaluating $\varphi$ at the monomial $z_{(a,x)}z_{(b,y)}$ would yield
\begin{align}
\varphi(z_{(a,x)}z_{(b,y)})&={\rm tr}\left(\rho \left( \left(q_A(a)(x)\otimes I_s\right)\cdot \left(I_d\otimes q_B(b)(y)\right)\right)\right)\\
&=p(x,y\mid a,b). \label{eq:last}
\end{align}
The crucial point is that  $\varphi$ evaluated at this monomial needs to have the value $p(x,y\mid a,b)$  \emph{for any strategy } realising $p$. In other words,  the linear constraint on $\varphi$ expressed in Equation \eqref{eq:last} must hold even if we do not know the strategy. 
This functional must satisfy other nice properties  independently of the strategy too, such as being positive. 

This perspective is precisely the one  we now take. Namely, we assume that the strategy $\rho, q_A, q_B$ is {\it not} given (since our question is whether $p$ is a quantum strategy at all), and 
we search for a functional on $\mathcal G$ that has the stated properties (or other properties, depending on the kind of strategies one  is looking for). 
When restricted to a finite-dimensional subspace of $\mathcal G$, this becomes a semidefinite optimisation problem, as can be easily checked. 
The dimension of this subspace will be the parameter indicating the level of the hierarchy, which is gradually increased. 
Solvability of all these semidefinite problems is thus a necessary condition for $p$ to be a quantum correlation matrix. 
In words, the levels of the NPA hierarchy form an outer approximation to the set of correlations. 
Conversely, if all/many of these problems are feasible, one can apply a (truncated) {\it  Gelfand-Naimark-Segal (GNS) construction} (see for example \cite{Pau02}) to the obtained functional, and thereby try to construct a  quantum strategy that realises $p$.
This is the content of the NPA hierarchy from the perspective of free semialgebraic geometry.

\subsection{Positivity in tensor networks}
\label{ssec:tn}

Let us finally explain some results about positivity in tensor networks. The results are not as much related to free semialgebraic geometry as   to positivity and sums of squares, as we will see. 

Since the state space of a composite quantum system is given by the tensor product of smaller state spaces (Eq.\ \eqref{eq:tensprod}), the global dimension $d$ grows exponentially with the number of subsystems $n$. 
Very soon it becomes infeasible to work with the entire space --- to describe $n=270$ qubits $d_i=2$, we would need to deal with a space dimension $d\sim 2^{270}\sim 10^{80}$, the estimated number of atoms in the Universe. 
 To describe anything at the macro-scale involving a mole of particles, $\sim10^{23}$, we would need a space dimension of  $\sim 2^{{10}^{23}}$, which is much larger than a googol ($10^{100}$), but smaller than a googolplex ($10^{10^{100}}$). 
 These absurd numbers illustrates how quickly the Hilbert space description becomes impractical --- 
in  practice, it works well for a few tens of qubits.\footnote{The lack of scalability of this description is far from being a unique case in physics --- most theories are not scalable. One needs to find the new relevant degrees of freedom at the new scale, which will define an emergent theory.}
 
Fortunately, many physically relevant states admit an efficient description. 
The ultimate reason is that physical interactions are local (w.r.t. a particular tensor product decomposition; this decomposition typically reflects  spatial locality). The resulting relevant states  admit a description using only a few terms for every local Hilbert space. The main idea of tensor networks is precisely to use a few matrices for every local Hilbert space ${\rm Mat}_{d_i}$ (Eq.\ \eqref{eq:tensprod}; see, e.g., \cite{Or18,Ci20}).

\begin{figure}[t]\centering
 \includegraphics[width=.5\columnwidth]{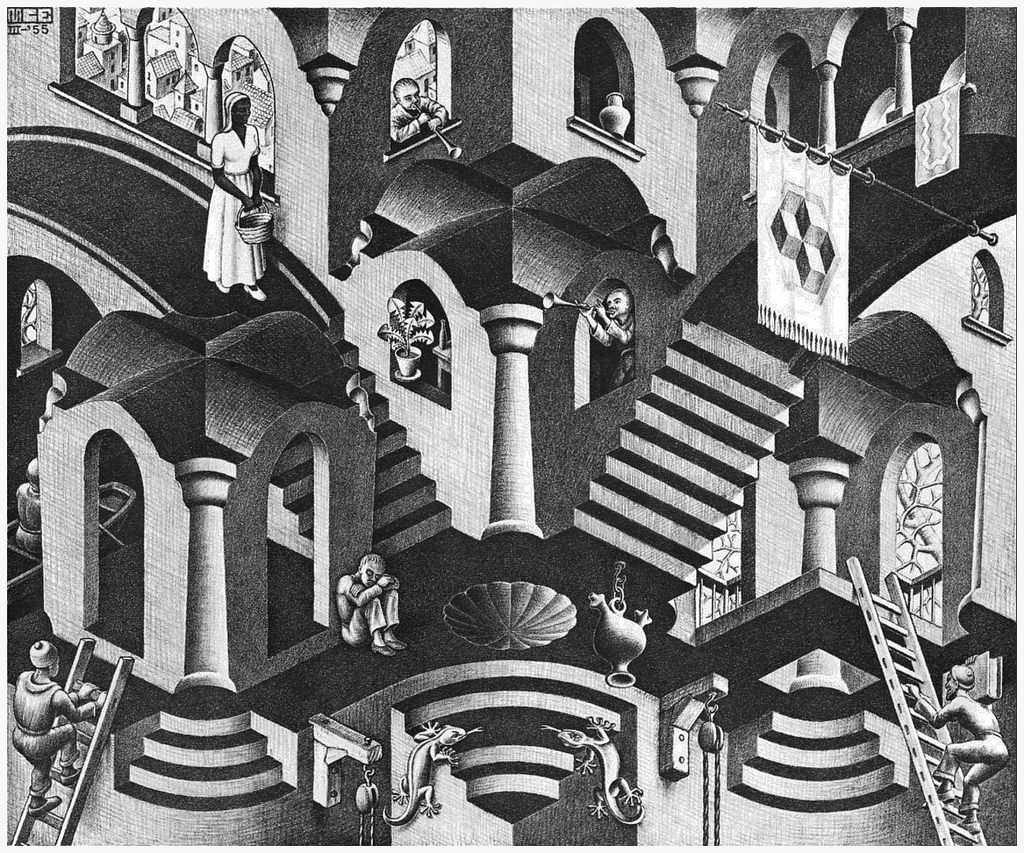}
 \caption{The notion of positivity gives rise to convexity, 
 which gives rise to many surprising effects when interacting with the multiplicity of systems, as in this lithograph by M.\ C.\ Escher. }
 \label{fig:convex}
 \end{figure}

Now, this idea interacts  with \emph{positivity} in a very interesting way (\cref{fig:convex}). 
Positivity  is a property in the global space ${\rm Mat}_d$ which cannot be easily translated  to positivity properties in the local spaces. 
As we will see, there is a `tension' between  using a few matrices for each local Hilbert space and representing the positivity locally. 
This mathematical interplay has implications for the description of quantum many-body systems, among others.  

Let us see one   example of a tensor network decomposition where this \emph{positivity problem} appears. 
To describe a mixed state in   one spatial dimension  with periodic boundary conditions we use 
the {\it matrix product density operator form} (MPDO) of $\rho$,  
$$
\rho=\sum_{i_1,\ldots, i_n=1}^r\rho_{i_1,i_2}^{(1)}\otimes \rho_{i_2,i_3}^{(2)}\otimes\cdots \otimes \rho_{i_n,i_1}^{(n)}. 
$$ 
The smallest such $r$ is called the {\it operator Schmidt rank} of $\rho$ \cite{Ve04b,Zw04}.  
Clearly, every state 
admits an MPDO form, and the ones with small $r$ can be handled efficiently. 
But how is the positivity of $\rho$ reflected in the local matrices?
Clearly, if all local  matrices are psd (i.e.\ $\rho_{ij}^{(k)}\succcurlyeq 0$) then $\rho$  will  be psd. 
But some sums of non-psd matrices will also give rise to a global psd matrix, since negative subspaces may cancel in the sum.
Can one easily characterise the set of local matrices whose sum is psd? The short answer is `no'. 

For further reference, if all local matrices are psd, so that $\rho$ is  separable, the corresponding $r$ is called the {\it separable rank }   of $\rho$ \cite{DN,DHN}.

To obtain a local certificate of positivity, we first express  $\rho=\xi\xi^*$ (which is possible only  if $\rho$ is psd) and then apply the tensor network `philosophy' to $\xi$, i.e.\ express $\xi$ as an MPDO:  
$$
\rho=\xi\xi^*\quad \mbox{ with }\quad \xi=\sum_{i_1,\ldots, i_n=1}^r\xi_{i_1,i_2}^{(1)}\otimes \xi_{i_2,i_3}^{(2)}\otimes\cdots \otimes \xi_{i_n,i_1}^{(n)}.
$$ 
This is the {\it local purification form} of $\rho$.
Note that there are many $\xi$ that satisfy $\rho=\xi\xi^*$, as $\xi$ need not be Hermitian or a square matrix (it could be a column vector).
The smallest  $r$ among all such $\xi$   is called the {\it purification rank} of $\rho$.

The interesting point for the purposes of this paper is that the purification rank is a \emph{noncommutative generalisation} of the positive semidefinite rank of a nonnegative matrix. 
There are many more such connections: the separable rank, 
the translational invariant (t.i.) purification rank, and 
the t.i. separable rank are noncommutative generalisations 
of the  nonnegative rank, the cpsd rank and the cp rank of nonnegative matrices, respectively \cite{DN}.
As a matter of fact, this connection holds in much greater generality, as we will explain below. 
In all of these cases, the ranks coincide for quantum states that are diagonal in the computational basis. 

From our perspective, this connection is beneficial for both sides. 
For example, for quantum many-body systems, this  insight together with the results by \cite{GPT} leads to the following result:

\begin{theorem}[\cite{DSPC,DN}] 
The purification rank cannot be upper bounded by a function of the operator Schmidt rank only.
The separable rank cannot be upper bounded by a function of the purification rank only.
\end{theorem}

(It is worth noting these separations are not robust, as they disappear in the approximate case for certain norms \cite{De20}.)

Conversely, the quantum perspective provides a natural and well-motivated path for generalisation of the `commutative' results about cpsd rank, cp rank, etc. 
For example, in \cite{GPT} it is shown that the extension complexity of a polytope w.r.t.\  a given cone   is given by the rank of the slack matrix of that polytope w.r.t.\ that cone. We wonder whether this result could be  generalisation to the noncommutative world. This would give a \emph{geometric} interpretation of the purification rank, the separable rank and their symmetric versions, perhaps as extension complexities of some objects.

\medskip
\begin{center}
\deco
\end{center}
\medskip

{\it Symmetry} is a central property  in physics, both conceptually and practically. Conceptually, symmetry is the other side of the coin of a conserved quantity (by Noether's theorem). Practically, it allows for more efficient mathematical descriptions, as symmetric objects have fewer degrees of freedom. 
For example, in the above context,   $\rho$ is  {\it translational invariant} if it remains unchanged under cyclic permutations of the local systems. 
This raises the question: is there an  MPDO form that explicitly expresses this  symmetry? For example, the following form does,  
$$
\rho=\sum_{i_1,\ldots, i_n=1}^r\rho_{i_1,i_2}\otimes \rho_{i_2,i_3}\otimes\cdots \otimes \rho_{i_n,i_1}, 
$$ 
because it uses the same matrices on every site, and the arrangement of indices is such that a cyclic permutation of the local systems does not change $\rho$. But does this hold for other symmetries too?

The existence of such \emph{invariant decompositions} and their  corresponding ranks has  been studied in a very general framework \cite{DHN}. 
Explicitly, every tensor decomposition is represented as a simplicial complex, 
where the individual tensor product spaces are associated to the vertices, and the summation indices to the facets. 
The symmetry is modelled by a group action on the simplicial complex. 
The central result is that an invariant decomposition exists if the group action is {\it free} on the simplicial complex \cite{DHN}. 
Just to give one example, if $\rho\in {\rm Mat}_d\otimes{\rm Mat}_d$ is separable and symmetric, 
it will in general {\it not} admit a decomposition of the type 
$$
\rho=\sum_\alpha \rho_\alpha\otimes\rho_\alpha \quad \mbox{with all } \rho_\alpha \mbox{ psd,}
$$ 
 but it will have one  of the type 
$$
\rho= \sum_{\alpha,\beta} \rho_{\alpha,\beta}\otimes\rho_{\beta,\alpha} \quad \mbox{with all } \rho_{\alpha,\beta} \mbox{ psd}.
$$ 
From the perspective of our framework, this is due the fact that the group permuting the two  end points of an edge  does not act freely on the edge connecting them. 
But this group action can be made free if the two points are connected by two edges, leading to the two indices $\alpha, \beta$ in the above sum.
This is one example of a \emph{refinement} of a simplicial complex, which makes the action of the group free \cite{DHN}. 

Finally, we remark that this framework of tensor decompositions with invariance can not only be applied to quantum-many body systems, but to any object in a tensor product space. 
One example are multivariate symmetric polynomials with positivity conditions \cite{Kl21}.

\medskip
\begin{center}
\deco
\end{center}
\medskip

A related question is the existence of invariant decompositions \emph{uniform} in the system size. 
Namely, given a tensor 
$\rho=\left(\rho_{\alpha,\beta}\right)_{\alpha,\beta=1,\ldots, r}$   with all  $\rho_{\alpha,\beta}\in {\rm Mat}_d$,
 define  
$$
\tau_n(\rho):=\sum_{i_1,\ldots, i_n=1}^r \rho_{\alpha_1,\alpha_2}\otimes \rho_{\alpha_2,\alpha_3}\otimes\cdots \otimes \rho_{\alpha_n,\alpha_1}\in {\rm Mat}_{d^n} 
$$
for all $n\in \N$. The result, in this case, is very different from the fixed 
$n$ case:

\begin{theorem}[\cite{DCCW}]\label{thm:undec}
Let $d,r\geqslant 7$. Then it is undecidable whether $\tau_n(\rho)\succcurlyeq 0$ for all $n\in\N$.
\end{theorem}

Using this result it can be shown that a translationally invariant local purification of $\tau_n(\rho)$ \emph{uniform in the system size} need not exist \cite{DCCW}.  

The proof of this theorem uses a reduction from the matrix mortality problem. 
In the latter,   given a finite set of matrices $M_\alpha\in {\rm Mat}_d(\mathbb{Z})$, one is asked whether there is a word $w$ such that $0=M_{w_1}\cdots M_{w_n} \in {\rm Mat}_{d}(\mathbb{Z})$. 
While this problem is noncommutative (because matrix multiplication is), 
the problem about $\tau_n(\rho)$ is `more' noncommutative. Intuitively, if all $\rho_{\alpha,\beta}$ are diagonal, we recover a version of the matrix mortality problem. Note also that the space where $\tau_n(\rho)$ lives grows with $n$, in contrast to the matrix mortality problem. 

The decidability of a similar problem can be studied for more general algebras. In that case, $\rho_{\alpha,\beta}$ is in a certain algebra, and $\tau_n(\rho)$ is asked to be in a certain cone  \cite{Gr21}.

\medskip
\begin{center}
\deco
\end{center}
\medskip

Let us finally explain a computational approach for the finite case.  
So consider $n$ fixed and recall that after specifying some local matrices $\rho_{i}^{(j)}$, one wants to know whether 
  $$
  \rho=\sum_i \rho_i^{(1)}\otimes \cdots \otimes \rho_i^{(n)}
  $$ 
is psd. Since the $n$ is fixed, this problem is decidable, but computing and diagonalising $\rho$   is impossible for large values of $n$ (in fact it is  \textsf{NP}-hard \cite{Kl14}). So one has to come up with a different idea. 
What   can  be computed are certain {\it moments} of $\rho$, i.e.\ the numbers $\tr(\rho^k)$ for small enough $k$. 
This follows from the observation that the moments only require local matrix products, 
$$
\tr(\rho^k)=\sum_{i_1,\ldots, i_k=1}^r \tr\left(\rho_{i_1}^{(1)}\cdots \rho_{i_k}^{(1)} \right)\cdots \tr\left(\rho_{i_1}^{(n)}\cdots \rho_{i_k}^{(n)} \right) .
$$
These few moments can then be used to compute optimal upper and lower bounds on the distance of $\rho$ to the cone of psd matrices \cite{D20/2}. Specifically, to compute this distance it suffices to compare $\rho$ with $f(\rho)$, 
 where $f\colon \R\to \R$ is the function that leaves the positive  numbers unchanged, and sets the negative  numbers  to zero. 
 We then approximate $f$ by polynomial functions $q$ of low degree, so that $\tr(q(\rho))$ only uses a few moments of $\rho$. 
The best results were obtained with certain {\it sums of squares approximations}, 
which can be computed with a linear or semidefinite optimisation.

\section{Closing words}
\label{sec:concl}

We have illustrated how quantum information theory and free semialgebraic geometry often study very similar mathematical objects  from different perspectives. 
We have given the examples of 
positivity and separability  (\cref{ssec:pos}), 
quantum magic squares (\cref{ssec:magic}), 
non-local games (\cref{ssec:game}), 
and positivity in tensor networks (\cref{ssec:tn}). 	
In all of these cases, we have tried to illustrate how results can be transferred among the two fields, and how this can be beneficial for the two perspectives. 
As mentioned in the introduction, there are many similar such connections which have not been covered here.

Going back to New Hampshire's motto, we conclude that it is undecidable to determine whether to live free or die, 
because both the question of whether matrices generate a free semigroup  
and 
the matrix mortality problem are undecidable. 

\bigskip
\emph{Acknowledgements}.---
GDLC acknowledges support from the Austrian Science Fund (FWF) with projects START Prize Y1261-N and the Stand Alone project P33122-N. 
TN acknowledges support from the Austrian Science Fund (FWF) with  Stand Alone project P29496-N35.


\end{document}